\documentclass[aps,showpacs]{revtex4}
\usepackage{graphicx}

\begin{document}

\title{Renormalization of the $\Phi^4$  scalar theory
under Robin boundary conditions and a possible new renormalization
ambiguity}
\author{Luiz C de Albuquerque$^{}$\footnote{\tt lclaudio@fatecsp.br }}
\affiliation{Faculdade de Tecnologia de S\~ao Paulo - CEETEPS - UNESP \\
Pra\c{c}a Fernando Prestes, 30, 01124-060 S\~ao Paulo, SP, Brazil}

\begin{abstract}
We perform a detailed analysis of renormalization at one-loop
order in the $\lambda\phi^4$ theory with Robin boundary condition
(characterized by a constant $c$) on a single plate at $z=0$. For
arbitrary $c\geq0$ the renormalized theory is finite after the
inclusion of the usual mass and coupling constant counterterms,
and two independent surface counterterms.
A surface counterterm renormalizes the parameter $c$.
The other one may involve either an additional
wave-function renormalization for fields at the surface, or an
extra quadratic surface counterterm. We show that both choices
lead to consistent subtraction schemes at one-loop order, and that
moreover it is possible to work out a consistent scheme with both
counterterms included. In this case, however, they can not be
independent quantities.  We study a simple one-parameter family of
solutions where they are assumed to be proportional to each other,
with a constant $\vartheta$. Moreover, we show that the
renormalized Green functions at one-loop order does not depend on
$\vartheta$. This result is interpreted as indicating a possible
new renormalization ambiguity related to the choice of
$\vartheta$.
\end{abstract}

\pacs{03.70.+k, 11.10.-z, 11.10.Kk, 11.10.Gh}
\maketitle

%\begin{center}
{\small {\it Keywords:} Boundary Conditions in QFT, Casimir Effect,
Perturbative Renormalization, Surface Critical Phenomena.}
%\end{center}

\section{Introduction}

Boundary conditions are of prime importance in the quantum domain.
Observables, i.e. Hermitian operators in a Hilbert space, are
defined only after its domain (as given by the set of boundary
condition) is specified. This observation underlies the
concept of self-adjoint extensions of an operator and
has far reaching consequences in quantum physics \cite{Bal,Asorey}.
For a recent application of these ideas in a
toy model for topology changing fluctuation in quantum gravity see
\cite{RandTopo}.

Boundary conditions (BC) are ubiquitous in QFT for several reasons.
A prototypical example is given by the Casimir effect,
i.e. an observable force produced by zero-point vacuum fluctuations
of a quantum field in the presence of boundaries or
uncharged surfaces \cite{Casimir}. In this case the
classical BC provides a simple mathematical model for
the otherwise complex interactions between the quantum field
and the classical body \cite{LC,Rey}. More recently
much attention has been paid to the question of  BC in
the context of the holographic principle and the brane-world scenario.

There is an ongoing effort to improve
the calculation  of the Casimir force arising
from the fluctuations of electromagnetic field.
This is necessary for a reliable  comparison with
recent high-precision measurements of the Casimir
force  between a flat plate and a spherical
surface (lens) or a sphere \cite{Exp1}, and also between two parallel
flat plates \cite{Exp2}. Moreover, the Casimir effect has
analogues in condensed matter physics, for instance in fluctuation
induced forces \cite{Kardar} and critical phenomena in
semi-confined systems \cite{Diehl}, and may even play a
key role in the design and operation of micro- and nano-scale
electromechanical devices \cite{Nano}.

A field $\phi$ is said to obey Robin boundary condition at a
surface $\Sigma$ if its normal derivative at a point on $\Sigma$
is proportional to its value there:
\begin {equation}\label{1}
\frac{ \partial }{ \partial n } \, \phi(x) = c\, \phi(x),
\qquad x \in \Sigma.
\end {equation}
Neumann and Dirichlet boundary conditions arise as particular cases
of Robin boundary condition: the first one corresponds to $c = 0$;
the other is obtained in the limit $c \to \infty$ (assuming that
$\partial_n \phi$ is bounded). In this paper we will restrict
ourselves to the case $c\geq0$.

The mixed case of Dirichlet-Robin  BC were considered in
\cite{Moste} for a 2D massless scalar field as a phenomenological
model for a penetrable surface, with $c^{-1}$ playing the role of the
finite penetration depth. Recently, the Casimir energy for a
scalar field subject to Robin BC on one or two parallel planes
(separated by a distance $a$) was computed in \cite{Romeo,JPA}.
There, it was shown that
for the mixed case of Dirichlet-Robin BC
the Casimir energy as a function of $a$ develops a minimum, i.e.,
there is a configuration of stable equilibrium for the position
of the planes (see also \cite{Bajnok1} for an interesting approach to the
computation of the Casimir energy).  Robin BC for scalar fields in the background 
of the Schwarzchild, de Sitter,
and conformally flat Brane-World geometries have been investigated
in \cite{Setare}, whereas the heat-kernel expansion for manifolds 
with Robin BC on the boundary was studied in \cite{Bordag}.
Moreover, Robin BC are relevant in
stabilization mechanisms of the compactification radius of large
extra dimensions in five and six dimensional orbifolds \cite{Pontom}.

In this paper we perform a detailed study of
renormalization  at one-loop order in the scalar $\lambda\phi^4$
in the presence of a flat surface at $z=0$ for Robin BC.
In contrast to the existing computations \cite{Diehl,Symanzik}
which pay attention to the particular cases $c=0,\infty$ and use a mixed
coordinate-momentum space regularization, we will keep $c\geq0$
arbitrary and workout the regularization entirely in momentum space.
This procedure avoids dealing with distributions
and test functions otherwise unavoidable in the coordinate space
regularization. Renormalization in the presence of hard boundaries \cite{LC}
poses new difficulties related to the loss of full Lorentz invariance.
Theories without Lorentz invariance raised a considerable theoretical
and experimental interest in recent times, see for instance \cite{NCFT,Vuce}.
This gives another motivation to our work, since again the BC leads to
a simple model to study renormalization in Lorentz non-invariant theories.

The present study continues the one-loop
renormalization which was started in \cite{JPA}, where
the connected two-point function at first order in $\lambda$
was considered. It was shown that in order to renormalize the two-point
Green function  one has to introduce
two independent surface counterterms at the same order
besides the usual (bulk, i.e without the surface at $z=0$)
mass counterterm at one-loop.
A surface counterterm renormalizes the parameter $c$,
and requires the introduction of a term of the form
$\delta c\delta(z)\phi^2$ in the counterterm Lagrangian. The other
demands the introduction of a new term
\begin{equation}\label{2}
\delta b\,\delta(z)\,\phi\,\left(\partial_z-c\right)\,\phi~.
\end{equation}
Here we will consider the connected four-point Green function at
second order in $\lambda$. In particular, it is shown that
the renormalized four-point function is finite with the inclusion of the
bulk coupling constant counterterm at order $\lambda^2$,
together with the one-loop bulk mass counterterm and the two
surface counterterms mentioned above. In other words,
no extra ultraviolet (UV) singularities in the four-point
function arises when the external points approach the surface.
This is in agreement with the general framework developed in
\cite{Diehl,Symanzik}.

Regarding the vertex (\ref{2}) there is a kind of ordering
ambiguity which appears when some or all external points
approach the surface at $z=0$ \cite{Diehl}.
In \cite{JPA} we followed
the prescription of first let the external points approach
the surface and later integrate over the vertex (\ref{2}).
This ordering will be called {\it AFIL} (\lq\lq Approach
First, Integrate Later''). On the other hand, it is possible
to proceed in the other way round, namely first integrate over
the vertex (\ref{2}) and later approach the surface \cite{Diehl}.
We call this prescription {\it IFAL} (\lq\lq Integrate First,
Approach Later''). The counterterm (\ref{2}) makes no contribution
at all with this ordering prescription since the factors
involving $\delta b$ cancel identically.
Thus, in this case one is forced to implement another subtraction
procedure. This is done by introducing an extra renormalization
constant $Z_s$ in order to allow for an additional renormalization
of the surface field \cite{Diehl}. We show that both
prescriptions lead to consistent subtraction schemes
at one-loop order. In fact, it is possible to go one step further.
Within the {\it AFIL} prescription, it is possible to include
both counterterms ($\delta\,b$ and $Z_s$). However, they cannot
be independent renormalization constants. We show that the
simple parameterization $Z_s=\vartheta\,\delta\, b$ leads to a
consistent renormalized theory at one-loop order. Moreover,
we prove that the renormalized Green functions do not depend on
the parameter $\vartheta$. Thus, the ordering ambiguity in dealing
with the vertex (\ref{2}) seems to lead to a new type
of renormalization ambiguity (involving the choice of $\vartheta$),
which was as far as we know unnoticed up to now.

The paper is organized as follows. In Section {\bf II} 
we review some relevant material and fix the notation \cite{JPA}.
In Section {\bf III} we revisit
the computation of the renormalized two-point Green function
at one-loop order, first  using the prescription {\it AFIL},
and then with the prescription {\it IFAL}. In the same Section
we investigate the choice $Z_s=\vartheta\,\delta\,b$.
Section {\bf IV} describes the renormalization of the
connected four-point function for the parameterization
$Z_s=\vartheta\,\delta\,b$. Finally, Section {\bf V}
is devoted to the conclusions and comments. Some useful formulas
are collected in three Appendixes.

%%%%%%%%%%%%%%%%%%%%%%%%%%%%%%%%%%%%%%%%%%%%%%%%%%%%%%%%%%%%%%%%%%%%%%%%%%%%
\section{Perturbation theory for Robin boundary condition}
\label{sec2}

Consider a real scalar field in $D=d+1$
dimensions living in the half-space $z\ge 0$ \footnote{Conventions:
$\hbar=c=1$, $x=({\bf x},z)$, ${\bf x}=(x^0,\ldots,x^{d-1})$.},
with the Euclidean action
\begin{equation}\label{3}
S[\phi]=\int d^d{\bf x}\,\int_0^{\infty}dz\left[\frac{1}{2}\,
(\partial_{\mu}\phi)^2
+U(\phi)+c\,\phi^2\,\delta(z)\right],
\end{equation}
The stationary action principle applied to the action (\ref{3})
gives \footnote{Notice that
$\int_0^{\infty}\delta(z)\,f(z)\,dz=\frac{1}{2}\,f(0)$.}
\begin{equation}\label{4}
\delta S=\int d^d{\bf x}\left\{(-\partial_z\phi+c\phi)\,\eta\,\Big|_{z=0}
+\int_0^{\infty}dz\,[-\partial^2\phi+U'(\phi)]\,\eta\right\}+O(\eta^2).
\end{equation}
From (\ref{4}) one gets the (Euclidean) equation of motion
\begin{equation}\label{5}
-\partial^2\phi+U'(\phi)=0
\end{equation}
and the Robin boundary condition at $z=0$,
\begin{equation}\label{6}
\partial_z\phi-c\phi\,\Big|_{z=0}=0.
\end{equation}
Furthermore, $\phi({\bf x},\infty)=0$ and $\phi({\bf x}\to\infty,z)=0$.

The partial Fourier transform of $\phi({\bf x},z)$ is
\begin{equation}\label{7}
\phi({\bf x},z)=\int\frac{d^d{\bf k}}{(2\pi)^d}\,e^{i{\bf k}\cdot{\bf x}}\,
\varphi({\bf k})\,e^{-kz},
\end{equation}
where $k=|{\bf k}|$. In this mixed $k-z$ representation,
the Feynman propagator (unperturbed two-point Green function) reads \cite{JPA}
(the dependence on $c$ will be omitted)
\begin{equation}\label{8}
{\cal G}({\bf k};z,z')={\cal G}^{\rm D}({\bf k};z,z')+
\delta{\cal G}({\bf k};z,z')
\end{equation}
where ($z_{<}(z_{>})={\rm min(max)}\{z,z'\}$)
\begin{equation}\label{9}
{\cal G}^{\rm D}({\bf k};z,z')=\frac{1}{k}\,\sinh(kz_<)\,\exp(-kz_>),
\end{equation}
is the Dirichlet propagator, ${\cal G}^{\rm D}({\bf k};0,z')=
{\cal G}^{\rm D}({\bf k};\infty,z')=0$, and
\begin{equation}\label{10}
\delta {\cal G}({\bf k};z,z')=\frac{e^{-k(z+z')}}{c+k},
\end{equation}
is a decreasing function of $c$. From (\ref{8})-(\ref{10}) it is possible
to obtain the alternative representation
\cite{Diehl}
\begin{equation}\label{11}
{\cal G}({\bf k};z,z')=\frac{1}{2k}\Bigl[~e^{-k\,|z -z^\prime|}
-\frac{c-k}{c+k}\,e^{-k\,(z+z')}~\Bigr],
\end{equation}
where the first term inside the square brackets is the bulk
(i.e. in the absence of the surface at $z=0$)
propagator of the scalar field. The Feynman propagator satisfies
the Robin BC at $z=0$,
\begin{equation}\label{12}
\partial_z\,{\cal G}({\bf k};z,z')\Bigr|_{z=0}=c\,{\cal G}({\bf k};0,z')~,
\end{equation}
and lacks full translational invariance due to the second term
inside the square brackets in Eq. (\ref{11}).

In the limit $z^\prime\to0$  one gets from (\ref{12})
\begin{equation}\label{13}
\lim_{z^\prime\to0}\,\Bigl[\,\lim_{z\to0}\,
\partial_z\,{\cal G}({\bf k};z,z')\Bigr]=c\,
{\cal G}({\bf k};0,0).
\end{equation}
There is an ordering ambiguity in the double limit $z\to0,
z^\prime\to0$ \cite{Diehl}, as can be seen by first letting
$z^\prime\to0$, and then taking the derivative at $z=0$,
\begin{equation}\label{14}
\lim_{z\to0}\,\Bigl[\,\lim_{z^\prime\to0}\,
\partial_z\,{\cal G}({\bf k};z,z')\Bigr]=-k\,
{\cal G}({\bf k};0,0).
\end{equation}

The implementation of BC via local terms in the action is
employed in  studies of boundary
critical phenomena \cite{Diehl}. In that context, 
it can be shown that Dirichlet
and Neumann BC correspond to the so-called ordinary ($c\to\infty$) and
special transitions ($c=0$), respectively. The Robin BC is relevant
in the study of the crossover between those universality classes,
for which, however, the computations become much more
involved. It is relevant also for the analysis of
the ordinary transition; in this case, however, one may resort to an
expansion in powers of $c^{-1}$ \cite{Diehl}.

It is convenient to distinguish between Green functions with
and  without points on the boundary. We then define the
Green functions
\begin{equation}\label{15}
G^{(N,L)}(x_1,...,x_N;{\bf x}_{N+1},...,{\bf x}_L):=
\langle\,\phi(x_1)...\phi(x_N)\phi_s({\bf x}_{N+1})...
\phi_s({\bf x}_{L})\,\rangle,
\end{equation}
where the notation is $\phi_s({\bf x})=\phi({\bf x},0)$.
Initially, however, we will work with $G^{(M)}(x_1,...,x_M)$
and later identify the Green functions $G^{(N,L)}$ by letting
$M-N=L$  external points approach the boundary.

In the $\lambda\phi^4$ theory
the only primitively UV divergent amplitudes are the
two- and four-point Green
functions\footnote{\footnotesize Composite operators will
not be considered in this paper.}. This property is not spoiled
by the Robin BC \cite{Symanzik,Diehl}.
In \cite{JPA} it was shown that to renormalize
the two-point Green function at first order in $\lambda$ it is 
necessary to include three renormalization constants ($\delta m^2$,
$\delta c$, and $\delta b$).
We will revisit this calculation, paying special attention
to the consequences of the ambiguity displayed in
(\ref{13}) and (\ref{14}).

The regularized Euclidean Lagrangian density of the theory
will be defined as
\begin{equation}\label{16}
{\cal L}_0(\phi_0;m_0,\lambda_0,c_0,b_0;\Lambda)=
\frac{1}{2}\,(\partial_{\mu}\phi_0)^2+\frac{1}{2}\,m_0^2
\phi_0^2+\frac{\lambda_0}{4!}\,\phi_0^4
+c_0\,\delta(z)\,\phi_0^2+b_0\,\delta(z)\,\phi_0\,\partial_z\phi_0,
\end{equation}
with the field $\phi_0$ living in the half-space $z\ge 0$, and
where the dependence on the cut-off scale $\Lambda$ is via
the bare parameters $m_0,~\lambda_0,~c_0$, $b_0$ and
bare field $\phi_0$.

The finite, renormalized Green functions are given from
\begin{equation}\label{17}
G_R^{(N,L)}(x_1,...,{\bf x}_L;m,\lambda,c,b;\kappa)=
\lim_{\Lambda\to\infty} Z^{-(N+L)/2}\,
G^{(N,L)}(x_1,...,{\bf x}_L;m_0,\lambda_0,c_0,b_0;\Lambda)
\end{equation}
with $G^{(N,L)}$ the regularized Green functions computed from
(\ref{16}) and $\kappa$ is a mass scale defining the subtraction point.
In the multiplicative renormalization scheme
\begin{eqnarray}\label{18}
&& m_0^2=Z^{-1}\,(m^2+\delta m^2)~,\\
&& \lambda_0=Z^{-2}\,(\lambda+\delta\lambda), \label{19}\\
&& c_0=Z^{-1}\,(c+\delta c)-Z^{-1}\,c\,( b-\delta b) \label{20}\\
&& b_0=Z^{-1}\,(b+\delta b),\label{21}
\end{eqnarray}
gives the relation between the renormalized and bare parameters.
The renormalized field
is $\phi=Z^{-1/2}\phi_0$, with $Z=1+\delta Z$. At $d=3$  $\delta m^2$
contains both a logarithmic and a quadratic divergence.
Thus it may be written as $\delta m^2= \delta Z_m m^2+\Delta m^2$.
In an analogous way, one may write $\delta c=\delta Z_c c+\Delta c$.
Both $\Delta m^2$ and $\Delta c$ vanish in the dimensional regularization
(DR) scheme.

In perturbation theory, the bare parameters and $Z$ in
Eqs. (\ref{18})-(\ref{21}) are written as power
series in $\lambda$ so that, for $\delta f\in\{\delta Z,~\delta m^2,
~\delta\lambda,~\delta c,~\delta b\}$,
\begin{equation}\label{22}
\delta f=\sum_{k=1}^\infty \lambda^k\,\delta f_k~,
\end{equation}
where the coefficients ($\delta Z_k,~\delta
m^2_k,~\delta\lambda_k,~\delta
c_k,~\delta b_k$) are functions of $m,~c,~b$ and $\Lambda$
(or $\epsilon$ in DR). Its singular part is chosen so that
the limit $\Lambda\to\infty$ ($\epsilon\to0$ in DR) in (\ref{17})
gives a well-defined function of the renormalized parameters and
$\kappa$. The finite part is fixed by the renormalization conditions (RC).

The mass, coupling constant, and wave-function counterterms can be
chosen as usual, namely using two conditions on the 1PI two-point
function and one condition on the 1PI four-point function, both
evaluated in the theory without the surface at $z=0$
\cite{Diehl,Symanzik}. Therefore, with this choice they take the same
values of the bulk theory for a given subtraction scheme.
In particular, $\delta Z_1=\delta\lambda_1=0$.
As for the surface renormalization constants, a natural choice
for RC is
\begin{eqnarray}\label{23}
&&{\cal G}^{(0,2)}_R({\bf k})\Bigr|_{k=\kappa}=(c+\kappa)^{-1},\\
&&\frac{d}{dk}{\cal G}^{(0,2)}_R({\bf k})\Bigr|_{k=\kappa}
=-(c+\kappa)^{-2}\label{24}
\end{eqnarray}
since they are satisfied at tree level, where
${\cal G}^{(0,2)}_R({\bf k})={\cal G}({\bf k};0,0)$. Notice that we will set
$m=0$. For this reason we have to be careful when $k=0$ is taken in
${\cal G}^{(0,2)}_R({\bf k})$ and its derivatives, 
specially in the particular case $c=0$. 
Moreover, one may employ a minimal subtraction scheme.

The defining relations (\ref{18})-(\ref{21}) can be enforced
directly on the Lagrangian density. We will study the particular case
$m=b=0$, so that
\begin{equation}\label{25}
{\cal L}_0(\phi_0;m=0,\lambda,c,b=0)={\cal L}^{(0)}
(\phi_0;c)+{\cal L}_{\rm I}(\phi_0;c,\lambda)~,
\end{equation}
where we have split ${\cal L}_0$ into a free, unperturbed part,
\begin{equation}\label{26}
{\cal L}^{(0)}(\phi_0;c)=
\frac{1}{2}\,(\partial_{\mu}\phi_0)^2+c\,\delta(z)\,\phi_0^2
\end{equation}
plus an interacting Lagrangian density, including counterterms.
It is sufficient to consider only the connected parts of
the two- and four-point Green functions.
The interacting Lagrangian density which gives the connected
parts of the two- and four-point Green functions
at one-loop order reduces to
\begin{equation}\label{28}
{\cal L}_{\rm I}(\phi_0;c,\lambda;\Lambda)=\frac{\lambda}{4!}\,
\phi_0^4+
\frac{\lambda}{2}\,\delta m_1^2\phi_0^2+
\frac{\lambda^2}{4!}\,\delta\lambda_2\phi_0^4
+\lambda\delta c_1\,\delta(z)\,\phi_0^2+
\lambda\delta b_1\,\delta(z)\,\phi_0\left(\partial_z-c\right)\phi_0
+O(\lambda^2)~.
\end{equation}
The only counterterm of order $\lambda^2$ omitted
in Eq. (\ref{28}) is $(Z^{-1}-1)c\,\delta(z)\,\phi_0^2$,
the others being of high-order in $\lambda$.
This is justified since it contributes to the two-point function
only at two-loops, and to a disconnected part of the four-point 
function at one-loop.

Notice that we have chosen not to enforce the relation
$\phi_0=Z^{1/2}\phi$ on the Lagrangian level. Thus there remains
an overall multiplicative renormalization, as displayed in Eq.
(\ref{17}).

\section{Two-point Green function at one-loop order}

The connected two-point Green function reads
\begin{eqnarray}\label{29}
&&{\cal G}^{(2)}({\bf k};z,z')=
{\cal G}({\bf k};z,z')+
\lambda\Biggl\{\delta{\cal G}_1({\bf k};z,z')-\delta m_1^2
\tilde{I}({\bf k};z,z')-
\delta c_1{\cal G}({\bf k};z,0){\cal G}({\bf k};0,z')\nonumber\\
&&\qquad\qquad\qquad-\delta b_1 \int_0^\infty dz''
\delta(z'')\Bigl[{\cal G}({\bf k};z,z'')
\left(\partial_{z''}-c\right){\cal G}({\bf k};z'',z')
+(z\leftrightarrow z')\Bigr]\Biggr\}+O(\lambda^2),
\end{eqnarray}
where
\begin{equation}\label{30}
\tilde{I}({\bf k};z,z'):=\int_0^\infty dz''\,{\cal G}({\bf k};z,z'')\,
{\cal G}({\bf k};z'',z'),
\end{equation}
and
\begin{equation}\label{31}
\delta{\cal G}_1({\bf k};z,z'):=-\frac{1}{2}\,\int\frac{d^d{\bf q}}
{(2\pi)^d}\,\int_0^\infty dz''\,{\cal G}({\bf k};z,z'')\,
{\cal G}({\bf q};z'',z'')\,{\cal G}({\bf k};z'',z')~.
\end{equation}
In Appendix {\bf A} we show that for $z>0,~z^\prime>0$ one may
expand $\delta{\cal G}_1({\bf k};z,z')$ as in Eq. (\ref{a9}),
\begin{equation}\label{32}
\delta{\cal G}_1({\bf k};z,z')= - I_0\,\tilde{I}({\bf k};z,z')
-J_0(0,c)\,{\cal G}({\bf k};z,0)\,{\cal G}({\bf k};0,z')+ 
\Delta_1{\cal G}_1({\bf k};z,z')~,
\end{equation}
where $\Delta_1{\cal G}_1({\bf k};z,z')$ is regular for $z>0,~z'>0$, and
$d<4$. Moreover,
\begin{eqnarray}\label{33}
&&J_n(k,c):=\frac{1}{8}\int \frac{d^d{\bf
q}}{(2\pi)^d}\, \frac{1}{q^n(q+c)(q+k)}~,\\
&&I_0:=\frac{1}{4}\int \frac{d^d{\bf
q}}{(2\pi)^d}\,\frac{1}{q}~.\label{34}
\end{eqnarray}
From (\ref{29})-(\ref{32}) we obtain
\begin{eqnarray}\label{35}
&&{\cal G}^{(2)}({\bf k};z,z')={\cal G}({\bf k};z,z')+
\lambda\,\Biggl\{\,\left(\,I_0+\delta m_1^2\,\right)\,
\tilde{I}({\bf k};z,z')-\left[\,J_0(0,c)+\delta c_1\,\right]\,
{\cal G}({\bf k};z,0)\,{\cal G}({\bf k};0,z')\,\,\nonumber\\
&&\qquad\quad+\Delta_1{\cal G}_1({\bf k};z,z') -
\delta b_1\int_0^\infty dz''
\delta(z'')\left[{\cal G}({\bf k};z,z'')
(\partial_{z''}-c){\cal G}({\bf k};z'',z')+(z\leftrightarrow
z')\right]\Biggr\} + O(\lambda^2).
\end{eqnarray}
For $z>0,~z'>0$ the last term in (\ref{35}) is zero due to the BC at
$z''=0$, see (\ref{12}). Moreover, adopting the RC 
($p^\mu=({\bf p},p_z)$ is a $d+1$ momentum)
\begin{equation}\label{36}
\Gamma_{R}^{(2)}(p^2=0)=0,
\end{equation}
on the 1PI two-point function of the bulk theory, one gets
\begin{equation}\label{37}
\delta m^2_1=- I_0~.
\end{equation}
Now we make the identification ${\cal G}^{(2,0)}({\bf k};z,z')=
{\cal G}^{(2)}({\bf k};z,z')$, and obtain from (\ref{17}) 
the consistency condition
\begin{equation}\label{38}
\lim_{\Lambda\to\infty}\,\left[\, J_0(0,c)+\delta c_1\,
\right]={\rm finite\,\,function\,\, for}\,\,d=3~.
\end{equation}
This condition must be satisfied at $O(\lambda)$.
Otherwise ${\cal G}^{(2,0)}_R$ will not be a regular function.

\subsection{{\it AFIL} prescription}

Now we will consider the limit $z\to0$ of 
$G^{(2)}({\bf k};z,z^\prime)$
with  $z'\neq0$. In this case, the ambiguity displayed in Eqs.
(\ref{13}) and (\ref{14}) will show up in the evaluation of
the last term on the RHS of (\ref{29}),
\begin{eqnarray}\label{40}
&&\lim_{z\to0}\int_0^\infty dz''\,
\delta(z'')\left[\,{\cal G}({\bf k};z,z'')
\left(\partial_{z''}-c\right){\cal G}({\bf k};z'',z')+
(z\leftrightarrow z')\,\right]\\
&&\qquad\qquad\qquad\qquad=-\frac{1}{2}\,
{\cal G}({\bf k};0,z'),\label{41}
\end{eqnarray}
where to obtain (\ref{41}) we followed the \lq\lq
ordering prescription''

{\it AFIL}:  First attach the external point to the boundary and later
integrate over the surface vertex.

This  choice was adopted in \cite{JPA}. For $z\to0$ and  $z'\neq0$,
the correct expansion of $\delta{\cal G}_1$ is given by Eq. (\ref{a12})
instead of Eq.(\ref{31}). From (\ref{29}), (\ref{a12}),
(\ref{37}), and (\ref{41}) we obtain for $z'>0$
\begin{eqnarray}\label{42}
\lim_{z\to0}\,{\cal G}^{(2)}({\bf k};z,z')&=&
{\cal G}({\bf k};0,z')-\lambda\Biggl\{\Bigl[\, J_0({\bf k},c)+
\frac{k-c}{2}\,J_1(k,c)-
\frac{c~(k+c)}{2}\,J_2(k,c)\nonumber\\
&&+\,\delta c_1-\frac{(c+k)}{2}\delta
b_1\,\Bigr]{\cal G}({\bf k};0,0)\,{\cal G}({\bf k};0,z')
-\Delta_2{\cal G}_1({\bf k};z')\Biggr\}+O(\lambda^2)~,
\end{eqnarray}
From now on we set ${\cal G}^{(1,1)}({\bf k};z')=
\lim_{z\to0}\,{\cal G}^{(2)}({\bf k};z,z')$.
From (\ref{17}) and (\ref{42}) one gets
\begin{equation}\label{43}
\lim_{\Lambda\to\infty}\,\left[\,
J_0(k,c)+\frac{k-c}{2}\,J_1(k,c)
+\delta c_1
-\frac{k+c}{2}\delta b_1\,
\right]={\rm finite\,\,fc.\,\, for\,\,} d=3~.
\end{equation}
If this condition is not satisfied the renormalized
${\cal G}^{(1,1)}_R({\bf k};z)$ will not be a regular function 
at $O(\lambda)$.

Now consider the limits $z\to0,~z'\to0$. This time, in place of
(\ref{32}) and (\ref{a12}) we will use Eq. (\ref{a15}).
Moreover, in the {\it AFIL} prescription one gets
\begin{equation}\label{45}
\lim_{z,z'\to0}\int_0^\infty dz''\,
\delta(z'')\left[\,{\cal G}({\bf k};z,z'')\,
\left(\partial_{z''}-c\right)\,{\cal G}({\bf k};z'',z')+
(z\leftrightarrow z')\,\right]\nonumber\\
=-\,{\cal G}({\bf k};0,0)~.
\end{equation}
With (\ref{a15}) and (\ref{45}), ${\cal G}^{(0,2)}_R({\bf k})$ reads
\begin{eqnarray}\label{46}
{\cal G}^{(0,2)}_R({\bf k})&&=\lim_{\Lambda\to\infty}
{\cal G}^{(2)}({\bf k};0,0)
= {\cal G}({\bf k};0,0)\nonumber\\
&&-\lambda\,\lim_{\Lambda\to\infty}\left[\, J_0(k,c)- cJ_1(k,c)
+\delta c_1-(c+k)\delta b_1\,\right]\,
\left[{\cal G}({\bf k};0,0)\right]^2+O(\lambda^2).
\end{eqnarray}
Imposing the RC (\ref{23}) and (\ref{24}) upon 
${\cal G}^{(0,2)}_R({\bf k})$
gives us  two equations whose solutions are
\begin{eqnarray}\label{49}
&&\delta b_1= J'_0(\kappa,c)-c J'_1(0,c)~,\\
&&\delta c_1=(c+\kappa)\left[\,
J'_0(\kappa,c)-c J'_1(\kappa,c)\right]-
J_0(\kappa,c)+c J_1(\kappa,c)~, \label{50}
\end{eqnarray}
where $J'_n(\kappa,c)=\partial J_n(k,c)/\partial k$ at $k=\kappa$.

Let us verify the consistency of this subtraction scheme.
Using (\ref{49}) and (\ref{50}) in (\ref{38}) and (\ref{43})
it is easy to show that
\begin{eqnarray}\label{51}
&&\lim_{\Lambda\to\infty}\,\left[\,J_0(0,c)+\delta c_1\,
\right]=-(c+\kappa)^2\,\lim_{\Lambda\to\infty}\,J'_1(\kappa,c),\\
&&\lim_{\Lambda\to\infty}\,\left[\,
 J_0(k,c)+\frac{k-c}{2}\,J_1(k,c)
+\delta c_1-\frac{k+c}{2}\delta b_1\,\right]\nonumber\\
&&\qquad\quad= -\frac{1}{2}\,\lim_{\Lambda\to\infty}\,\left[
k(k+c)\,J_2(k,c)+(c+\kappa)(c+2\kappa)\,J_1^\prime(\kappa,c)+
\kappa^2(c+\kappa)\,J_2^\prime(\kappa,c)\,\right]~,\label{52}
\end{eqnarray}
are  regular functions for $d<4$. We conclude that the choice (\ref{37}),
(\ref{49}) and (\ref{50}) for $\delta m^2_1$, $\delta b_1$ and
$\delta c_1$, respectively, lead to a well defined one-loop two-point
Green function.

\subsection{{\it IFAL} prescription}

There is no a priori reason to adopt the ordering prescription
{\it AFIL}. Instead it is quite conceivable to take the opposite route
\cite{Diehl,Diehl2}, namely:

{\it IFAL}: First compute the integral over the surface vertex and
later let the external point approach  the surface.

In this case, using (\ref{13}) we obtain in place of (\ref{41})
\begin{equation}\label{53}
\lim_{z\to0}\int_0^\infty dz''\,
\delta(z'')\left[\,{\cal G}({\bf k};z,z'')
\left(\partial_{z''}-c\right){\cal G}({\bf k};z'',z')+
(z\leftrightarrow z')\,\right]=0~.
\end{equation}
We conclude that the boundary
counterterm (\ref{2}) is ineffective in the {\it IFAL} prescription,
and we may set $\delta b=0$ from the start.
However, Eqs. (\ref{43}) and (\ref{50})
cannot be satisfied for $\delta b=0$.

In order to obtain the finite, renormalized Green functions
one has to introduce an extra renormalization constant $Z_s$,
to allow for an independent renormalization of the surface
field \cite{Diehl},
\begin{equation}\label{54}
\phi_0\vert_s=Z^{1/2}Z_s^{1/2}\phi({\bf x},0)
=Z_s^{1/2}\phi_0({\bf x},0),
\end{equation}
with $Z_s=1+\delta Z_s$. Now, instead of
(\ref{17}) one defines (with $b=b_0=0$)
\begin{equation}\label{55}
G_R^{(N,L)}(x_1,...,{\bf x}_L;m,\lambda,c;\kappa)=
\lim_{\Lambda\to\infty} Z^{-(N+L)/2}Z_s^{-L/2}
G^{(N,L)}(x_1,...,{\bf x}_L;m_0,\lambda_0,c_0;\Lambda)
\end{equation}
with the multiplicative renormalization
\begin{equation}\label{56}
c_0=Z^{-1}Z_s^{-1}(c+\delta c)
\end{equation}
replacing (\ref{20}). Moreover, with (\ref{54}) and (\ref{56}) the vertex
$\lambda\delta c_1\delta(z)\phi_0^2$ in Eq. (\ref{28}) is replaced by
\begin{equation}\label{56b}
\lambda\left(\delta c_1-c\delta Z_{s,1}\right)\delta(z)\phi_0^2~.
\end{equation}
Notice that since $\delta Z_s\neq0$
it is not only convenient but also necessary to distinguish
between  Green functions with and without $L=0$.
If we repeat the previous analysis
we will end up with two conditions and two equations to be satisfied.
The two equations are solved to give
\begin{eqnarray}\label{57}
&&\delta Z_{s,1}= -J'_0(\kappa,c)+c J'_1(0,c)~,\\
&&\delta c_1= \kappa\left[\,
J'_0(\kappa,c)-c J'_1(\kappa,c)\right]-
J_0(\kappa,c)+c J_1(\kappa,c)~,\label{58}
\end{eqnarray}
whereas the two consistency conditions are now
\begin{eqnarray}\label{59}
&&\lim_{\Lambda\to\infty}\,\left[\,J_0(0,c)+\delta c_1-c\,\delta
  Z_{s,1}\,\right]={\rm finite\,\,function\,\, for}\,\,d=3~,\\
&&\lim_{\Lambda\to\infty}\left[\,J_0(k,c)+
\frac{k-c}{2}J_1(k,c)
+\delta c_1+\frac{k-c}{2}\delta Z_{s,1}\,
\right]={\rm finite\,\,fc.\,\, for}\,\,d=3~,\label{60}
\end{eqnarray}
Using Eqs.(\ref{57}) and (\ref{58}) one verifies that
they are given by the RHS of Eqs.(\ref{51}) and (\ref{52}), respectively.
We conclude that the renormalized two-point Green function
at one-loop do not depend on the choice of the ordering prescription.

\subsection{Mixed subtraction scheme: {\it AFIL}}

Within the {\it AFIL} prescription it is
possible to go one step further by introducing both counterterms,
$\delta b$ and $\delta Z_{s}$. In this case, with the definition
(\ref{54}), Eqs. (\ref{20}) and (\ref{21}) are replaced by ($b=0$)
\begin{eqnarray}\label{61}
&& c_0=Z^{-1}Z_s^{-1}(c+\delta c-c\delta b)\\
&& b_0=Z^{-1}Z_s^{-1}\delta b.~\label{62}
\end{eqnarray}
Moreover,
\begin{equation}\label{63}
G_R^{(N,L)}(x_1,...,{\bf x}_L;m,\lambda,c,b;\kappa)=
\lim_{\Lambda\to\infty} Z^{-(N+L)/2}\,Z_s^{-L/2}\,
G^{(N,L)}(x_1,...,{\bf x}_L;m_0,\lambda_0,c_0,b_0;\Lambda)
\end{equation}
is the generalization of (\ref{17}) and (\ref{55}). Repeating the
previous computations, we obtain that
\begin{eqnarray}\label{64}
&&\lim_{\Lambda\to\infty}\,\left[\, J_0(0,c)+\delta c_1-c\delta
  Z_{s,1}\,\right]~,\\
&&\lim_{\Lambda\to\infty}\left[\, J_0(k,c)+
\frac{k-c}{2}J_1(k,c)
+\delta c_1+\frac{k-c}{2}\delta Z_{s,1}-\frac{k+c}{2}\delta b_1\,
\right]~,\label{65}
\end{eqnarray}
must be finite functions at $d=3$. In addition, 
one gets the two equations
\begin{eqnarray}\label{66}
&&J_0(\kappa,c)- c J_1(\kappa,c)+
\delta c_1-(c+\kappa)\delta b_1+\kappa\delta Z_{s,1}=0~,\\
&&\delta b_1-\delta Z_{s,1}
-J'_0(\kappa,c)+c J'_1(\kappa,c)=0~,\label{67}
\end{eqnarray}
It is clear that $\delta b_1$
and $\delta Z_{s,1}$ can not be independent constants, since only
two conditions, (\ref{23}) and (\ref{24}), are available. In other
words, they are redundant renormalization constants. Let us assume
that
\begin{equation}\label{68}
\delta Z_s=\vartheta\delta b.
\end{equation}
with $\vartheta\ne1$ ($\vartheta=1$ is ruled out by (\ref{67})).
The solutions to (\ref{66}), (\ref{67}) and (\ref{68}) can be
cast in the form
\begin{eqnarray}\label{69}
&&\delta c_1=\left(\kappa+\frac{c}{1-\vartheta}\right)\,
\left[\,J'_0(\kappa,c;\Lambda)-c J'_1(\kappa,c;\Lambda)\right]-
J_0(\kappa,c;\Lambda)+c J_1(\kappa,c;\Lambda)~,\\
&&\delta b_1=\frac{1}{1-\vartheta}\left[\,J'_0(\kappa,c;\Lambda)-
c J'_1(\kappa,c;\Lambda)\right]~,\label{70}\\
&&\delta
Z_{s,1}=\frac{\vartheta}{1-\vartheta}\,\left[\,J'_0(\kappa,c;\Lambda)-
c J'_1(\kappa,c;\Lambda)\right]~,\label{71}
\end{eqnarray}
where we have explicitly indicated the dependence of $J_0$ and
$J_1$ on the cut-off $\Lambda$, see Appendix {\bf B}. The surface
counterterms are functions of $\vartheta$. Renormalized quantities
however can not depend on the choice of $\vartheta$. Indeed, using
(\ref{69}) and (\ref{70}) one shows from (\ref{64}) and (\ref{65})
that they are given again by the RHS of (\ref{51}) and (\ref{52}),
respectively. The choice $\delta Z_{s}=0$ is parameterized by the
limit $\vartheta\to0$, and leads to the results (\ref{49}) and
(\ref{50}). On the other hand, taking $\vartheta\to\infty$
reproduces (\ref{57}) and (\ref{58}), corresponding to the case
$\delta b=0$ in the {\it IFAL} prescription.

Using the results in Appendix {\bf B} the counterterms 
may be written as ($d=3$)
\begin{eqnarray}\label{75}
&&\delta c_1=-\frac{1}{16\pi^2}\left[\,\Lambda
-c\,\frac{k^2(1-2\vartheta)+
c\,(c+2\kappa)}{(1-\vartheta)\,(c-\kappa)^2}
\,{\rm ln}\left(\frac{\Lambda}{\kappa}\right) ~+~
2c^2\,\frac{2\kappa\,(1-\vartheta)+c\vartheta}
{(1-\vartheta)\,(c-\kappa)^2}
\,{\rm ln}\left(\frac{\Lambda}{c}\right)\,\right]~+~
\delta\bar{c}_1~,\\
&&\delta b_1=-\frac{1}{16\pi^2}\frac{1}{1-\vartheta}\,
\frac{1}{(c-\kappa)^2}
\left[\,(\kappa^2-2c\,\kappa-c^2)\,{\rm ln}\left(\frac{\Lambda}
{\kappa}\right)~
+~2c^2\,{\rm ln}\left(\frac{\Lambda}{c}\right)\right]~+~
\delta\bar{b}_1~,\label{76}\\
&&\delta
Z_{s,1}=-\frac{1}{16\pi^2}\frac{\vartheta}{1-\vartheta}\,
\frac{1}{(c-\kappa)^2}
\left[\,(\kappa^2-2c\,\kappa-c^2)\,{\rm ln}\left(\frac{\Lambda}
{\kappa}\right)~
+~2c^2\,{\rm ln}\left(\frac{\Lambda}{c}\right)\right]~+~
\delta\bar{Z}_{s,1}~,\label{77}
\end{eqnarray}
where $\delta\bar{b}_1$, $\delta\bar{c}_1$, and 
$\delta \bar{Z}_{s,1}$ are finite contributions.

In \cite{Diehl2} the surface counterterms were computed at second
order in $\lambda$
for the particular case $c=0$ in the DR scheme.
Moreover the {\it IFAL} prescription was implicitly
assumed. This corresponds to the limit $\vartheta\to\infty$ of Eqs.
(\ref{75})-(\ref{77}).
Notice that the counterterm $\delta Z_{s}$ corresponds to
$\delta Z_1$ in \cite{Diehl2}. In the limit $\vartheta\to\infty$ and $c=0$
we get from (\ref{77}) (only the divergent part is displayed)
\begin{equation}\label{e1}
Z_{s}=1~+~\frac{\lambda}{16\pi^2}\,{\rm ln}
\left(\frac{\Lambda}{\kappa}\right)~+~O(\lambda^2)~,
\end{equation}
This agrees with the result quoted in \cite{Diehl2} after identifying
${\rm ln}(\Lambda/\kappa)$ with $\epsilon^{-1}$. 
From (\ref{61}) we get for
$c_0$ (with $\delta c=c\,\delta Z_c+\Delta c$,
$Z_c=1+\delta Z_c$, and $\delta b=0$ for $\vartheta\to\infty$)
\begin{equation}\label{e2}
c_0=Z^{-1}\,Z_s^{-1}\,Z_c\,c~+~Z^{-1}\,Z_s^{-1}\,\Delta c~,
\end{equation}
Our definition of $Z_c$ is thus related to the analogous
definition in \cite{Diehl2} by
\begin{equation}\label{e3}
Z^{\rm D}_c=Z^{-1}\,Z_s^{-1}\,Z_c~.
\end{equation}
From (\ref{75}) and (\ref{77}) we obtain in the limit 
$\vartheta\to\infty$ and $c=0$ (divergent part only)
\begin{equation}\label{e4}
Z^{\rm D}_c=1~+~\frac{\lambda}{16\pi^2}\,{\rm ln}
\left(\frac{\Lambda}{\kappa}\right)~+~O(\lambda^2)~,
\end{equation}
which is the result quoted in \cite{Diehl2}
(${\rm ln}(\Lambda/\kappa)\to\epsilon^{-1}$).

A remarkable feature of (\ref{75}) is that the logarithmic divergent
part of $\delta c_1$ depends on $\vartheta$.
In particular, for $c\to0$ we have
\begin{equation}\label{e5}
Z_c=1~-~\frac{\lambda}{16\pi^2}\,
\left(\frac{2\vartheta-1}{1-\vartheta}\right)\,
{\rm ln}\left(\frac{\Lambda}{\kappa}\right)~+~O(\lambda^2)~,
\end{equation}
plus regular $O(\lambda)$ corrections. Therefore, in this case 
the choice $\vartheta=1/2$ renders $Z_c$ finite to one-loop order.
Linear divergences are absent in the DR scheme, implying that 
$\Delta c =0$.
Hence in the DR scheme it seems possible to tune $\vartheta$
so as to eliminate {\it all} divergent
terms from $\delta c_1$. Renormalization in this case is
implemented by $\delta b_1$ and $\delta Z_{s,1}$.

Now the case $c\to\infty$, i.e. a Dirichlet BC on $z=0$. 
It seems from (\ref{75})-(\ref{77}) that the limit $c\to\infty$ 
is ill defined. However this is not true. Using the formulas 
given above it is possible to show that, in the case $c\to\infty$,
the connected two-point Green function satisfies the Dirichlet BC
at tree level and at one-loop order. Indeed, the 
Dirichlet BC is preserved at each order in perturbation theory 
\cite{Symanzik}. Therefore from (\ref{35}),
(\ref{42}),  and (\ref{46}), for instance, it is easy to 
see that {\it the surface divergences are harmless in the 
Dirichlet case}, and there is no reason
whatsoever to introduce the surface counterterms. 
This can only happens for $c\to\infty$.
For general $c$ the renormalized two-point Green 
function {\it does not} satisfy
the Robin BC at $z=0$ because of the surface counterterms.

To conclude this Section, we remark that there is a kind of
ambiguity in the possible choice (parameterized by $\vartheta$) of
$\delta Z_s$, $\delta b$, and $\delta c$. This ambiguity 
can not be fixed at first order in $\lambda$, unless a 
particular choice of the ordering is adopted in the evaluation 
of (\ref{40}) (e.g. namely, the {\it IFAL} prescription ). 
This is the procedure followed in studies of surface critical 
phenomena \cite{Diehl,Diehl2}. Despite this,
one may argue that the physical renormalized Green functions do
not depend on the choice of $\vartheta$. This has been verified
explicitly at one-loop for the two point Green function. In the
next Section we make a first step towards a high-order
verification of this conjecture by studying the four-point Green
function at the second order in $\lambda$.

\section{Four-Point Green Function at One-Loop Order}

The partial Fourier transformed  four-point Green 
function may be written as
\begin{equation}\label{78}
{\cal G}^{(4)}({\bf p}_1,{\bf p}_2,{\bf p}_3,{\bf p}_4;
z_1,z_2,z_3,z_4)=
(2\pi)^d\,\delta^{(d)}\Bigl(\sum_{j=1}^4 {\bf p}_j\Bigr)\,
\tilde{\cal G}^{(4)}(\{{\bf p}_\ell\},\{z_\ell\}),
\end{equation}
where $\{{\bf p}_\ell\}$ is a shorthand for 
${\bf p}_1,{\bf p}_2,{\bf p}_3,{\bf p}_4$, and analogously
for $\{z_\ell\}$. In the same notation, the 
connected four-point function at first order in 0 $\lambda$ is
\begin{equation}\label{79}
\tilde{\cal G}^{(4) [1]}(\{{\bf p}_\ell\},\{z_\ell\})=-\lambda\,
\int_0^\infty dz\, {\cal G}({\bf p}_1,z_1)\,{\cal G}({\bf p}_2,z_2)\,
{\cal G}({\bf p}_3,z_3)\,{\cal G}({\bf p}_4,z_4),
\end{equation}
The second order corrections to the connected four-point function
are shown in Figure (1). The one-loop amplitude given 
in Fig. (1a) reads
\begin{eqnarray}\label{80}
\tilde{\cal G}^{({\bf s})}(\{{\bf p}_\ell\},\{z_\ell\})&=
&\frac{\lambda^2}{2}\,\int_0^\infty dz dz^\prime\prod_{\ell=1}^2
{\cal G}({\bf p}_\ell;z_\ell,z)\,\prod_{j=3}^4
{\cal G}({\bf p}_j;z^\prime,z_j)\nonumber\\
&&\;\;\times\int \frac{d^d{\bf p}}{(2\pi)^d}\,
{\cal G}({\bf p};z,z^\prime)\,{\cal G}({\bf p}-{\bf s};z^\prime,z)~,
\end{eqnarray}
where ${\bf s}={\bf p}_1+{\bf p}_2$. Diagrams (1b) and (1c) 
may be read off from (\ref{80}) by exchanging 
$({\bf p}_2,z_2)\leftrightarrow ({\bf p}_3,z_3)$, 
${\bf s}\to{\bf t}={\bf p}_1+{\bf p}_3$,
and $({\bf p}_2,z_2)\leftrightarrow ({\bf p}_4,z_4)$, 
${\bf s}\to{\bf u}={\bf p}_1+{\bf p}_4$, respectively.

\subsection{Proof that $\tilde{\cal G}^{({\bf s})}
(\{{\bf p}_\ell\},\{z_\ell\})$ does not contain surface singularities}

In the following we will show that the amplitudes given in
Figs. (1a), (1b), and (1c)
contain only bulk UV divergent terms. In other words,
$\tilde{\cal G}^{(\beta)}(\{{\bf p}_\ell\},\{z_\ell\})$ 
($\beta={\bf s},{\bf t},{\bf u}$)
does not develop UV divergences when any or all $z_\ell$ 
approach the surface at
$z=0$. To prove that it is enough to consider the case
$z_1=z_2=z_3=z_4=0$. Eq. (\ref{80}) may be
written as a sum of four terms,
\begin{equation}\label{81}
\tilde{\cal G}^{({\bf s})}(\{{\bf p}_\ell\},\{0\})=
\frac{\lambda^2}{2}\,\sum_{k=1}^4\,
\tilde{\cal G}^{({\bf s})}_j(\{{\bf p}_\ell\},\{0\})~,
\end{equation}
where the integrand of $\tilde{\cal G}^{({\bf s})}_1$, 
$\tilde{\cal G}^{({\bf s})}_2$,
$\tilde{\cal G}^{({\bf s})}_3$, and $\tilde{\cal G}^{({\bf s})}_4$ 
contains the factor $e^{-p|z-z^\prime|-p^\prime(z+z^\prime)}$,
$e^{-p^\prime|z-z^\prime|-p(z+z^\prime)}$,
$e^{-(p+p^\prime)(z+z^\prime)}$, and
$e^{-(p+p^\prime)|z-z^\prime|}$, respectively,
and $p^\prime=|{\bf p}-{\bf s}|$.

After the integrals over $z$ and $z^\prime$ are performed
$\tilde{\cal G}^{({\bf s})}_1(\{{\bf p}_\ell\},\{0\})$ reads
\begin{eqnarray}\label{83}
& &\tilde{\cal G}_1^{({\bf s})}(\{{\bf p}_\ell\},\{0\})=-\frac{1}{4}\,
\prod_{\ell=1}^4
\frac{1}{p_\ell+c}\,\int\frac{d^d{\bf p}}{(2\pi)^d}\,
\frac{1}{p~|{\bf p}-{\bf s}|}\,
\left(\frac{c-|{\bf p}-{\bf s}|}{c+|{\bf p}-{\bf s}|}
\right)\nonumber\\
& &\quad\times\frac{2(p+|{\bf p}-{\bf s}|)+\sum_{k=1}^4 p_k}
{(p+|{\bf p}-{\bf s}|+p_1+p_2)(p+|{\bf p}-{\bf s}|+p_3+p_4)
(2 |{\bf p}-{\bf s}|+\sum_{k=1}^4 p_k)}
\end{eqnarray}
This term will give an UV finite contribution to
$\tilde{\cal G}^{({\bf s})}(\{{\bf p}_\ell\},\{0\})$ for
$d<4$. The term $\tilde{\cal G}_2^{({\bf s})}(\{{\bf p}_\ell\},\{0\})$
in Eq. (\ref{81}) can be obtained from Eq. (\ref{83}) by the
exchange $p\leftrightarrow |{\bf p}-{\bf s}|$,
\begin{eqnarray}\label{84}
& &\tilde{\cal G}_2^{({\bf s})}(\{{\bf p}_\ell\},\{0\})=-
\frac{1}{4}\,\prod_{\ell=1}^4
\frac{1}{p_\ell+c}\,\int\frac{d^d{\bf p}}{(2\pi)^d}\,
\frac{1}{p~|{\bf p}-{\bf s}|}\,\left(\frac{c-p}{c+p}
\right)\nonumber\\
& &\quad\times\frac{2(p+|{\bf p}-{\bf s}|)+\sum_{k=1}^4 p_k}
{(p+|{\bf p}-{\bf s}|+p_1+p_2)(p+|{\bf p}-{\bf s}|+p_3+p_4)
(2p+\sum_{k=1}^4 p_k)}
\end{eqnarray}
In a similar way
\begin{eqnarray}\label{85}
\tilde{\cal G}_3^{({\bf s})}(\{{\bf p}_\ell\},\{0\})
&=&\frac{1}{4}\,\prod_{\ell=1}^4
\frac{1}{p_\ell+c}\,\int\frac{d^d{\bf p}}{(2\pi)^d}\,
\frac{1}{p~|{\bf p}-{\bf s}|}\,\left(\frac{c-p}{c+p}\right)\,
\left(\frac{c-|{\bf p}-{\bf s}|}{c+|{\bf p}-{\bf s}|}
\right)\nonumber\\
& &\times\frac{1}{(p+|{\bf p}-{\bf s}|+p_1+p_2)
(p+|{\bf p}-{\bf s}|+p_3+p_4)}
\end{eqnarray}
This term also gives a regular contribution to
$\tilde{\cal G}^{({\bf s})}(\{{\bf p}_\ell\},\{0\})$
at $d<4$ for non-exceptional external momenta ${\bf s}\ne0$.
In the case ${\bf s}=0$ it reduces to
\begin{equation}\label{86}
\tilde{\cal G}_3^{({\bf s}=0)}(\{{\bf p}_\ell\},\{0\})=
\frac{1}{16}\,\prod_{\ell=1}^4
\frac{1}{p_\ell+c}\,\int\frac{d^d{\bf p}}{(2\pi)^d}\,
\frac{1}{p^2\,(p+p_1)^2}\,\left(\frac{c-p}{c+p}\right)^2,
\end{equation}
which is UV finite for $d<4$ as well as IR finite for $p_1\ne0$.
Notice that there is a possible IR problem for $p_1=0$. This
however has nothing to do with the surface at $z=0$, since it also
occurs in the bulk theory in the massless case considered here.
Similar remarks applies to $\tilde{\cal G}_1^{({\bf s}=0)}$ and
$\tilde{\cal G}_2^{({\bf s}=0)}$.
Finally, the last contribution to
$\tilde{\cal G}^{({\bf s})}(\{{\bf p}_\ell\},\{0\})$ 
in Eq. (\ref{81}) reads
\begin{eqnarray}\label{87}
& &\tilde{\cal G}_4^{({\bf s})}(\{{\bf p}_\ell\},\{0\})=
\frac{1}{4\sum_{k=1}^4 p_k}\prod_{\ell=1}^4
\frac{1}{p_\ell+c}\int\frac{d^d{\bf p}}{(2\pi)^d}\,
\frac{1}{p~|{\bf p}-{\bf s}|}\left[
\frac{1}{p+|{\bf p}-{\bf s}|+p_1+p_2}+
\frac{1}{p+|{\bf p}-{\bf s}|+p_3+p_4}\right] .
\end{eqnarray}
This term contains an UV divergent part for $d\geq3$. Noticing
that
\begin{equation}\label{88}
\int_0^\infty dz \prod_{\ell=1}^4 {\cal G}({\bf p}_\ell;0,z)=
\frac{1}{4\sum_{k=1}^4 p_k}\,\prod_{\ell=1}^4
\frac{1}{p_\ell+c}~,
\end{equation}
and defining
\begin{equation}\label{89}
\Lambda_{{\bf s}}:=\frac{1}{8}\, \,\int\frac{d^d{\bf p}}{(2\pi)^d}\,
\frac{1}{p~|{\bf p}-{\bf s}|}\,\left[\,\frac{1}
{p+|{\bf p}-{\bf s}|+p_1+p_2}+
\frac{1}{p+|{\bf p}-{\bf s}|+p_3+p_4}\,\right]~,
\end{equation}
one may write Eq. (\ref{87}) as
\begin{equation}\label{90}
\tilde{\cal G}_4^{({\bf s})}(\{{\bf p}_\ell\},\{0\})=
\Lambda_{{\bf s}}\,\int_0^\infty dz \prod_{\ell=1}^4 
{\cal G}({\bf p}_\ell;0,z)~.
\end{equation}
Altogether, Eq. (\ref{81}) may be cast in the form
\begin{equation}\label{91}
\tilde{\cal G}^{({\bf s})}(\{{\bf p}_\ell\},\{0\})=
\lambda^2\left(\,
\Lambda_{{\bf s}}+\delta\Lambda_{{\bf s}}\,\right)\, 
\int_0^\infty dz\prod_{\ell=1}^4 {\cal G}({\bf p}_\ell;0,z)~,
\end{equation}
where $\delta\Lambda_{{\bf s}}$ contains the regular ($d<4$)
contributions given in Eqs. (\ref{83})-(\ref{85}).

The other \lq\lq channels" (${\bf t}$ and ${\bf u}$) give similar results.
Therefore
\begin{equation}\label{92}
\sum_{\beta={\bf s},{\bf t},{\bf u}}
\tilde{\cal G}^{(\beta)}(\{{\bf p}_\ell\},\{0\})
=\lambda^2\Biggl\{\,\sum_{\beta={\bf s},{\bf t},{\bf u}}\,
\Lambda_{\beta}+
\sum_{\beta={\bf s},{\bf t},{\bf u}}\,\delta\Lambda_{\beta}\,
\Biggr\}\,
\int_0^\infty dz \prod_{\ell=1}^4 {\cal G}({\bf p}_\ell;0,z)~,
\end{equation}
The UV divergences are localized in the contributions given by
$\Lambda_\beta$ ($\beta={\bf s},~{\bf t},~{\bf u}$). 
They are of logarithmic
nature at $d=3$. It is clear that they are cancelled by the
contribution coming from Fig. (1d), which is of the same form,
\begin{equation}\label{93}
{\cal G}_{\rm
d}(\{{\bf p}_\ell\},\{0\})=-\lambda^2\,
\delta\lambda_2\,\int_0^\infty dz
\prod_{\ell=1}^4 {\cal G}({\bf p}_\ell;0,z)~.
\end{equation}
Now, in Appendix {\bf C} is shown that the analogue of Eq.
(\ref{92}) for the case without the surface at $z=0$ is (see Eq.
(\ref{b8}))
\begin{equation}\label{94}
\sum_{\beta={\bf s},{\bf t},{\bf u}}
\tilde{\cal G}_B^{(\beta)}(\{{\bf p}_\ell\},\{0\})=\lambda^2
\Biggl\{\,\sum_{\beta={\bf s},{\bf t},{\bf u}}\,\Lambda_{\beta}+
\sum_{\beta={\bf s},{\bf t},{\bf u}}\,\delta
\Lambda_{\beta,B}\,\Biggr\}\,
\int_0^\infty dz \prod_{\ell=1}^4 {\cal G}_B({\bf p}_\ell;0,z)~,
\end{equation}
with {\it the same} $\Lambda_{\beta}$ given above, and a finite
($d<4$) contribution $\delta\Lambda_{\beta,B}$.

Therefore, since the divergent part of Eqs. (\ref{94}) (in the
absence of the surface) and (\ref{92}) (surface at $z=0$) are
exactly equal, we conclude that (i) There are no surface
divergences in $\tilde{\cal G}^{(\beta)}(\{{\bf p}_\ell\},\{0\})$
($\beta={\bf s},{\bf t},{\bf u}$); and (ii) 
The counterterm $\delta\lambda_2$
in Eq. (\ref{93}) may be chosen equal to the one-loop coupling
constant counterterm $\delta\lambda_2$ of the bulk theory.
Moreover, the above procedure can be repeated for any number of
external points $z_{\ell}$ outside of the surface at $z=0$. The
conclusions are  the same, the only difference being in the
regular terms (for $d<4$) $\delta\Lambda_\beta$ in Eq. (\ref{92}).

\subsection{The other one-loop amplitudes}

In the previous section we studied graphs (a)-(d) in Fig.(1). To
complete the discussion, we will discuss the contributions
associated with one-loop corrections to the external legs of
${\cal G}^{(4)[1]}$ in Eq. (\ref{79}). These are given by the diagrams
(e)-(t) in Fig.(1). In the following we will briefly discuss the
case $z_1=z_2=z_3=z_4=0$. The case with some or all external
points outside of the surface can be analyzed in the same way.
Moreover, we will work with the mixed subtraction scheme of
section {\bf III.C} since it allows the inclusion of both $\delta
b$ and $\delta Z_s$. For $z_1=z_2=z_3=z_4=0$, graphs (e)-(h)  read
\begin{eqnarray}\label{95}
&&\tilde{{\cal G}}_{\rm e}(\{{\bf p}_\ell\},\{0\})=
-\lambda\int_0^\infty dz \,\delta{\cal G}_1({\bf p}_1;0,z)\,
\prod_{\ell=2}^4 {\cal G}({\bf p}_\ell;z,0)~,\\
&&
\tilde{{\cal G}}_{\rm f}(\{{\bf p}_\ell\},\{0\})=\lambda^2\,
\delta m_1^2\, \int_0^\infty dz^\prime
\,{\cal G}({\bf p}_1;0,z^\prime)\int_0^\infty dz\, 
{\cal G}({\bf p}_1;z^\prime,z)\,
\prod_{\ell=2}^4
{\cal G}({\bf p}_\ell;z,0)~,\label{98}\\
&&
\tilde{{\cal G}}_{\rm g}(\{{\bf p}_\ell\},\{0\})=
2\lambda^2\Delta c_1
\int_0^\infty dz^\prime \,\delta(z^\prime)
{\cal G}({\bf p}_1;0,z^\prime)
\int_0^\infty dz \,{\cal G}({\bf p}_1;z^\prime,z)
\prod_{\ell=2}^4 {\cal G}({\bf p}_\ell;z,0)
~,\label{99}\\
&&\tilde{{\cal G}}_{\rm h}(\{{\bf p}_\ell\},\{0\})=2\lambda^2
\delta b_1\,\int_0^\infty dz\,\prod_{\ell=2}^4\,
{\cal G}({\bf p}_\ell;z,0)\int_0^\infty dz^\prime \,
\delta(z^\prime)\,{\cal G}({\bf p}_1;0,z^\prime)
\,\left(\partial_{z^\prime}-c\right)\,
{\cal G}({\bf p}_1;z^\prime,z)~.\label{100}
\end{eqnarray}
where $\Delta c_1=\delta c_1 -c \delta Z_{s,1}$.
The contributions from graphs
(i)-(l), (m)-(p), and (q)-(t) may be obtained from
(\ref{95})-(\ref{100}) by the replacements
${\bf p}_1\leftrightarrow {\bf p}_2$,
${\bf p}_1\leftrightarrow {\bf p}_3$, and
${\bf p}_1\leftrightarrow {\bf p}_4$, respectively.

It is known that for Robin BC the 1PI functions are not 
multiplicatively renormalized. This means that 1PR graphs must 
be included in a modified skeleton expansion \cite{Diehl,Symanzik}.
However, the connected Green functions are 
multiplicatively renormalized. We have shown that the 
amplitudes given by graphs (a)-(c) contain only bulk UV
divergences. Thus surface divergences in the four-point function 
may arise only in connection with graph (e). Clearly no new 
surface divergences are introduced beside the ones discussed 
in Section {\bf III}. These divergences are eliminated
by the one-loop counterterm insertions in graphs (g) and (h). 
This has been verified using the results from Section {\bf III}.

In this way one obtains the renormalized
four-point function at one-loop,
\begin{eqnarray}\label{102}
\tilde{{\cal G}}^{(0,4)}_R(\{{\bf p}_\ell\})&=&
\lim_{\Lambda\to\infty}\,
\left(1+\lambda\delta Z_{s,1}+
O(\lambda^2)\right)^{-2}\,\Biggl[\,\tilde{{\cal G}}^{(0,4)[1]}
(\{{\bf p}_\ell\})+\sum_{\beta={\bf s},{\bf t},{\bf u}}
\tilde{\cal G}^{(\beta)}(\{{\bf p}_\ell\},\{0\})\nonumber\\
&&\quad\quad+\sum_{\alpha={\rm e,...,u}}
\tilde{{\cal G}}_{\alpha}(\{{\bf p}_\ell\},\{0\})+
O(\lambda^3)\,\Biggr]\nonumber\\
&=&\tilde{{\cal G}}^{(0,4)[1]}(\{{\bf p}_\ell\})+
\lim_{\Lambda\to\infty}\,\lambda^2\,\Biggl[\,
\sum_{\beta={\bf s},{\bf t},{\bf u}}\Lambda_\beta-
\delta\lambda_2\,\Biggr]\,\int_0^\infty dz\,
\prod_{\ell=1}^4 {\cal G}({\bf p}_\ell;z,0)\nonumber\\
&&+\lim_{\Lambda\to\infty}\,\lambda^2\,\sum_{k=1}^4\biggl[\,
\delta c_1-c\delta Z_{s,1}-c\delta b_1+J_0(p_k,c)+
\frac{p_k-c}{2}\left(\,J_1(p_k,c)-\delta b_1\,\right)
\nonumber\\
&&+\frac{1}{2}(c+p_k)\delta Z_{s,1}\,\biggr]\,
{\cal G}({\bf p}_k;0,0)\,\int_0^\infty dz\,
\prod_{\ell=1}^4 {\cal G}({\bf p}_\ell;z,0)+ 
\Delta\tilde{{\cal G}}^{(0,4)}(\{{\bf p}_\ell\})+O(\lambda^3)~,
\end{eqnarray}
where $\Delta\tilde{{\cal G}}^{(0,4)}(\{{\bf p}_\ell\})$ 
includes $O(\lambda^2)$ regular contributions for $d<4$. 
As we have shown in section {\bf IV.A},
the logarithmic (at $d=3$) UV divergence in $\Lambda_\beta$
can be eliminated by a convenient choice of RC in the bulk theory.
Therefore, the second term on the RHS of (\ref{102})
is finite in the limit $\Lambda\to\infty$. As for the third  term
on the RHS, the following condition has to be met ($k=1,...,4$)
\begin{equation}\label{103}
\lim_{\Lambda\to\infty}\,\biggl[\, \delta
c_1-c\delta Z_{s,1}-c\delta
b_1+J_0(p_k,c)+\frac{p_k-c}{2}\left(\,J_1(p_k,c)
-\delta b_1\,\right) +\frac{1}{2}(c+p_k)\delta
Z_{s,1}\,\biggr] ={\rm finite}~,
\end{equation}
for $d<4$. This is the equation given in (\ref{65}), 
and it is satisfied as
shown in section {\bf III.C}, with
$\delta Z_{s,1}$, $\delta b_1$ and $\delta c_1$
given in Eqs. (\ref{69})-(\ref{71}).

We conclude that the
renormalized connected four-point function does not depend on the
choice of $\vartheta$, since the RHS of (\ref{103}) has no
dependence on $\vartheta$ as shown in  section {\bf III.C}.
Therefore, ${\cal G}^{(0,4)}(\{{\bf p}_\ell\})$ is finite up to
$O(\lambda^3)$ with the choice of {\it four} independent
counterterms: a set of two independent surface counterterms,
\begin{equation}\label{104}
\{\delta c_1(\vartheta),\delta b_1(\vartheta), 
\delta Z_{s,1}(\vartheta)\},
\quad {\rm for}\,\, \vartheta\neq1\nonumber
\end{equation}
in addition to the two bulk counterterms $\delta m_1^2$ 
and $\delta\lambda_2$.

Finally we remark that the inclusion of another flat surface at
$z=a$ does not change the overall picture presented here. Suppose
that at $z=a$ the field satisfies a Robin BC parameterized by
another constant $\tilde{c}$. Surface divergences localized on
$z=a$ arise, but they are of the same type of the ones appearing
in the case of a single flat surface at $z=0$. Therefore the new
set of surface counterterms associated to the surface at $z=a$ is
given by Eqs. (\ref{69})-(\ref{71}) upon the replacement
$c\to\tilde{c}$. In other words, surface counterterms of a new
type are not required.

%%%%%%%%%%%%%%%%%%%%%%%%%%%%%%%%%%%%%%%%%%%%%%%%%%%%%%%%%%%%%%%%%%

\section{Conclusions}

Boundary conditions are a key ingredient in  quantum theory.
It can be shown for instance that not all BC are consistent with the
conservation of probability in quantum mechanics \cite{Bal,Asorey}. 
Possible BC include periodic, Dirichlet, Neumann, Robin, and other BC
without a classical interpretation. Boundary conditions play a 
major role in the Casimir effect. 
The presence of surfaces or boundaries in the vacuum
leads to a manifestation of the universal zero-point fluctuations 
intrinsic to any quantum system in the form of the Casimir force.

Renormalization theory in the presence of surfaces is hampered 
by difficulties associated with the loss of full Lorentz invariance. 
However, a number of results
as well as a general framework may be developed for some BC 
\cite{Symanzik,Diehl}. 
In this paper we performed a detailed renormalization analysis  of
the $\lambda\phi^4$ theory at one-loop order in the case of Robin BC
at a single surface at $z=0$. In contrast to previous investigations
\cite{Symanzik,Diehl} we keep the parameter $c\geq 0$ arbitrary and
workout the regularization entirely in momentum space. Moreover, 
attention has been paid to the consequences of the ambiguity in 
the choice of the ordering prescription when dealing with 
insertions of the counterterm vertex (\ref{2}).

As far as we know, the prescription of first integrate over 
the surface vertex and then let the external point approach 
the surface ({\it IFAL} prescription) is always tacitly assumed 
in the existing literature \cite{Diehl,Diehl2}.
In this case, the surface vertex (\ref{2}) makes no contribution
at all, and one is forced to introduce another
counterterm related to an extra renormalization of the surface fields, 
see Eq.(\ref{54}). However, we argued that the other prescription, 
e.g. first attach the external point to the surface and then 
integrate over the vertex ({\it AFIL} prescription), also
leads to a well defined renormalized theory at one-loop order. 
In fact, working with the {\it AFIL} prescription it is possible 
to include both counterterms $\delta b$ and $\delta Z_s$. 
They cannot of course be independent quantities.
We have investigated a one-parameter family of solutions where 
$\delta Z_s = \vartheta \delta b$, and found that it leads to a 
consistent subtraction scheme at one-loop order. Moreover, 
we have shown that the renormalized two- and four-point connected 
Green functions do not depend on the choice of $\vartheta$ at 
one-loop order. We interpret this result as indicating a 
possible new renormalization ambiguity,
related to the choice of $\vartheta$.

Indeed, since the renormalized Green functions do not depend on
$\vartheta$ one immediately obtains the equation
\begin{eqnarray}\label{105}
&&\frac{d}{d\vartheta}\,{\cal G}_R^{(N,L)}
\left(\{{\bf p}_{\ell}\},\{z_{\ell}\}\right) 
\,=\, 0\nonumber\\
&&\qquad =\lim_{\Lambda\to\infty}\,\frac{d}{d\vartheta}\,
Z^{-\frac{N+L}{2}}\,Z_s^{-\frac{L}{2}}\,
{\cal G}^{(N,L)}\left(\{{\bf p}_{\ell}\},\{z_{\ell}\};m_0,
\lambda_0,c_0,b_0;\Lambda\right)~,
\end{eqnarray}
where on the RHS the bare Green functions are computed from
(\ref{16}). For instance, assuming that the bare Green functions 
depend on $\vartheta$ only through the surface counterterms, 
one obtains for the case $N=0$, $L=2$ a self-consistency 
condition,
\begin{equation}\label{106}
\left(\,\sigma_c\,\frac{\partial}{\partial c_0}+\sigma_b\,
\frac{\partial}{\partial b_0}\right)\,
{\cal G}^{(0,2)}({\bf p})=\sigma_s\,{\cal G}^{(0,2)}({\bf p})~,
\end{equation}
where 
\begin{eqnarray}\label{107}
&&\sigma_c\equiv \frac{\partial c_0}{\partial\vartheta}~;\nonumber\\
&&\sigma_b\equiv \frac{\partial b_0}{\partial\vartheta}
=-\frac{1}{c}\,\sigma_c+O\left(\lambda^2\right)~;\nonumber\\
&&\sigma_s\equiv \frac{\partial{\rm ln}\,Z_s}{\partial\vartheta}
=-\frac{1}{c}\,\sigma_c+O\left(\lambda^2\right)~.
\end{eqnarray}
It is easy to show that condition (\ref{106}) is  satisfied at 
one-loop order. On the other hand, it is possible to turn
(\ref{105}) into a renormalization-group-like equation, for instance
assuming that the bare Green functions bear an explicit
dependence on the parameter $\vartheta$.

It is conceivable however that a high-order calculation
may rule out some values of $\vartheta$ as inconsistent.
This remains to be verified, and for this reason we intend to carry
our computation of the surface counterterms to the next order
of perturbation theory.
We are also investigating the application
of the framework developed here to some problems in surface critical 
phenomena, in particular to the crossover between the ordinary and
the special universality classes. It seems worth to explore
Eq. (\ref{105}) in this context. 

Note: After the completion of this work we became aware of the paper \cite{Bajnok2} 
where a different Lagrangian quantization scheme for boundary quantum field theory
is developed.  There, the authors start from the Neumann BC and then define the
more general interacting theory  as a perturbation around this free theory.
They also show how the Robin BC can be obtained in this way. It would be interesting
to find the relation between the approach pursued here and the one in \cite{Bajnok2}.

%%%%%%%%%%%%%%%%%%%%%%%%%%%%%%%%%%%%%%%%%%%%%%%%%%%%%%%%%%%%%%%%%%%%%%%%
\acknowledgments
I thank Ricardo M. Cavalcanti for many illuminating discussions,
and for participation in early stages of this work, and
Adilson J. da Silva and Marcelo Gomes for a critical reading
of the paper. I would like to thank the Mathematical Physics
Department of USP at S\~ao Paulo for their kind hospitality.
This work was partially supported by CNPq, grant 307843/2003-3.
%%%%%%%%%%%%%%%%%%%%%%%%%%%%%%%%%%%%%%%%%%%%%%%%%%%%%%%%%%%%%%%%%%

\bigskip\bigskip
\renewcommand{\theequation}{A.\arabic{equation}}
\setcounter{equation}{0}
{\bf APPENDIX A}
\bigskip

Let us define
\begin{equation}\label{a1}
I({\bf k},{\bf q};z,z^\prime):=\int_0^\infty dw\,e^{-2qw}\,
{\cal G}_0({\bf k};z,w)\,{\cal G}_0({\bf k};w,z').
\end{equation}
Notice that $I({\bf k},{\bf q};z,z^\prime)= I({\bf k},{\bf q};z^\prime,z)$.
Using the propagator (\ref{11}) and doing the integral over $w$ 
one obtains,
\begin{equation}\label{a2}
I({\bf k},{\bf q};z,z^\prime)=I_1({\bf k},{\bf q};z,z^\prime)+
I_2({\bf k},{\bf q};z,z^\prime)~,
\end{equation}
where
\begin{eqnarray}\label{a3}
&&I_1({\bf k},{\bf q};z,z^\prime)=\frac{1}{2(2k)^2}\Biggl\{
\,\left(\frac{c-k}{c+k}\right)\,\frac{k}{q(q+k)}\,
e^{-k(z+z^\prime)-2qz}\\
&&\quad
+\,\theta(z-z^\prime)\,e^{-k(z-z^\prime)}\,
\Biggl[\,\frac{1}{q}\,
\left(e^{-2qz^\prime}-e^{-2qz}\right)+\left(\frac{e^{-2qz}}{q+k}-
\frac{e^{-2qz^\prime}}{q-k}\right)\,\Biggr]\Biggr\}
+(z\leftrightarrow z^\prime)~,\nonumber
\end{eqnarray}
and
\begin{equation}\label{a4}
I_2({\bf k},{\bf q};z,z^\prime)=\frac{1}{(2k)^2}\,
\frac{e^{-k(z+z^\prime)}}{(k+c)^2}\,
\Bigl[\,\frac{(c+k)^2}{2(q-k)}+\frac{(c-k)^2}{2(q+k)}-
\frac{(c+k)(c-k)}{q}\,\Bigr]
\end{equation}
Both $I_1$ in Eq. (\ref{a3}) and $I_2$ in Eq. (\ref{a4}) 
are symmetric functions of $z,~z^\prime$.

From the definition of $\delta{\cal G}_1({\bf k};z,z^\prime)$ 
in Eq. (\ref{31}) one gets
\begin{eqnarray}\label{a5}
\delta{\cal G}_1({\bf k};z,z^\prime)&=&-I_0\,
\tilde{I}({\bf k};z,z^\prime)-
\frac{1}{2}\int\frac{d^d{\bf q}}{(2\pi)^d}\,
\left(\,\frac{1}{c+q}-\frac{1}{2q}\right)\,
I({\bf k},{\bf q};z,z^\prime)\nonumber\\
&=&-I_0\,\tilde{I}({\bf k};z,z^\prime)-
\frac{1}{2}\int\frac{d^d{\bf q}}{(2\pi)^d}\,
\left(\,\frac{1}{c+q}-\frac{1}{2q}\right)\,
I_1({\bf k},{\bf q};z,z^\prime)\nonumber\\
&&-\frac{1}{2}\int\frac{d^d{\bf q}}{(2\pi)^d}\,
\left(\,\frac{1}{c+q}-\frac{1}{2q}\right)\,
I_2({\bf k},{\bf q};z,z^\prime)~,
\end{eqnarray}
where $\tilde{I}({\bf k};z,z^\prime)$ and $I_0$ are 
defined in Eqs. (\ref{30}) and (\ref{34}),
respectively. Now, it is clear that due to the exponential terms
$e^{-qz}$ and $e^{-qz^\prime}$ in $I_1({\bf k},{\bf q};z,z^\prime)$,
the integral over ${\bf q}$ in the second term of the 
RHS of (\ref{a5}) is convergent, unless $z=0$ and/or $z^\prime=0$. 
Assume for the moment that $z>0$ and $z^\prime>0$. 
Thus, the only divergent contributions to 
$\delta{\cal G}_1({\bf k};z,z^\prime)$ comes from the term 
containing $I_2({\bf k},{\bf q};z,z^\prime)$,
apart from the first term on the RHS of (\ref{a5}) 
which gives the (usual) bulk correction to the one-loop self-energy.

In order to get the leading divergent terms in the third term 
of (\ref{a5}) let us expand $I_2({\bf k},{\bf q};z,z^\prime)$. 
For large $q$ one obtains
\begin{equation}\label{a6}
I_2({\bf k},{\bf q};z,z^\prime)\,\stackrel{q\to\infty}{=}\,\frac{e^{-
k(z+z')}}{(c+k)^2}\left[\frac{1}{2q}+\frac{c}{2q^2}+
O\left(\frac{1}{q^3}\right)\right].
\end{equation}
Using (\ref{a6}) inside the integrand of the third term on the 
RHS of (\ref{a5}) one gets its behavior for large ${\bf q}$,
\begin{eqnarray}\label{a7}
&&-\frac{1}{2}\int\frac{d^d{\bf q}}{(2\pi)^d}\,
\left(\,\frac{1}{c+q}-\frac{1}{2q}\right)\,
I_2({\bf k},{\bf q};z,z^\prime)\,\sim\,
-\frac{1}{2}\frac{e^{-k(z+z')}}{(c+k)^2}
\int\frac{d^d{\bf q}}{(2\pi)^d}\,
\left[\frac{1}{4q^2}-\frac{c}{4q^3}+O\left(\frac{1}{q^4}\right)
\right]\nonumber\\
&&\quad\quad\sim\,
-\frac{1}{2}\frac{e^{-k(z+z')}}{(c+k)^2}\int\frac{d^d{\bf q}}
{(2\pi)^d}\,\left[\frac{1}{4q(q+c)}+O\left(\frac{1}{q^4}
\right)\right]\nonumber\\
&&\quad\quad=-J_0(0,c)\,{\cal G}({\bf k};z,0)\,
{\cal G}({\bf k};0,z')+
({\rm regular~~terms~~in~~the~~UV})~,
\end{eqnarray}
where $J_0(k;c)$ was defined in Eq. (\ref{33}). This means that
\begin{equation}\label{a8}
\overline{\Delta}{\cal G}_1({\bf k};z,z^\prime):=-
\frac{1}{2}\int\frac{d^d{\bf q}}{(2\pi)^d}\,
\left(\,\frac{1}{c+q}-\frac{1}{2q}\right)\,
I_2({\bf k},{\bf q};z,z^\prime)+J_0(0,c)\,{\cal G}({\bf k};z,0)\,
{\cal G}({\bf k};0,z')~,
\end{equation}
is a regular function for $d<4$ and {\it any value} of 
$z,~z^\prime$. Finally, we write Eq. (\ref{a5}) as
\begin{equation}\label{a9} 
\delta{\cal G}_1({\bf k};z,z^\prime)=-I_0\,
\tilde{I}({\bf k};z,z^\prime)-
J_0(0,c)\,{\cal G}({\bf k};z,0)\,{\cal G}({\bf k};0,z')+
\Delta_1{\cal G}_1({\bf k};z,z^\prime)~, 
\end{equation}
where
\begin{equation}\label{a10}
\Delta_1{\cal G}_1({\bf k};z,z^\prime)=-
\frac{1}{2}\int\frac{d^d{\bf q}}{(2\pi)^d}\,
\left(\,\frac{1}{c+q}-\frac{1}{2q}\right)\,I_1({\bf k},{\bf q};z,z^\prime)+
\overline{\Delta}{\cal G}_1({\bf k};z,z^\prime)~, 
\end{equation}
is a regular function for $z>0$, $z^\prime>0$, and $d<4$.

Consider now the case $z=0$ and $z^\prime>0$. From Eqs. 
(\ref{a2})-(\ref{a4}) we obtain
\begin{equation}\label{a11}
I({\bf k},{\bf q};0,z^\prime)=\frac{1}{4}\left[\,
\frac{c+k+2q}{q(q+k)}-\frac{c+k}{q(q+k)}\,
e^{-2qz^\prime}\,\right]\,{\cal G}({\bf k};0,0)\,
{\cal G}({\bf k};0,z')~.
\end{equation}
Using Eq. (\ref{a11}) in Eq.(\ref{31}) leads after 
some manipulations to the result
\begin{equation}\label{a12}
\delta{\cal G}_1({\bf k};0,z')=
-\left[\,\frac{1}{2k}I_0 + J_0(k,c)+\frac{k-c}{2}\,J_1(k,c)-
\frac{c~(k+c)}{2}\,J_2(k,c)\,\right]{\cal G}({\bf k};0,0)\,
{\cal G}({\bf k};0,z')+\Delta_2{\cal G}_1({\bf k};z')~,
\end{equation}
where
\begin{equation}\label{a13}
\Delta_2{\cal G}_1({\bf k};z')=\frac{c+k}{8}\int\frac{d^d{\bf
q}}{(2\pi)^d}\,
e^{-2qz^\prime}\,\frac{1}{q(q+k)}\,\left(\,\frac{1}{c+q}-
\frac{1}{2q}\right)\,
{\cal G}({\bf k};0,0)\,{\cal G}({\bf k};0,z')
\end{equation}
gives a regular contribution for
$z^\prime\neq0$.

Finally, consider the case $z=0$ and $z^\prime=0$. Again, 
from Eqs. (\ref{a2})-(\ref{a4}) one gets
\begin{equation}\label{a14}
I({\bf k},{\bf q};0,0)=\frac{1}{2(q+k)}\,\left[{\cal G}({\bf k};0,0)
\right]^2~.
\end{equation}
Substituting Eq. (\ref{a14}) in Eq. (\ref{31}) leads to
\begin{equation}\label{a15}
\delta{\cal G}_1({\bf k};0,0)=
-\left[\,\frac{1}{2k}I_0+J_0(k,c)-cJ_1(k,c)\,\right] \,
\left[\,{\cal G}({\bf k};0,0)\,\right]^2.
\end{equation}
\bigskip

%%%%%%%%%%%%%%%%%%%%%%%%%%%%%%%%%%%%%%%%%%%%%%%%%%%%%%%%%%%%%%%%%%%%%%%%%%%%
\renewcommand{\theequation}{B.\arabic{equation}}
\setcounter{equation}{0}
{\bf APPENDIX B}
\bigskip

Here we discuss the computation of
\begin{equation}\label{d1}
J_n(k,c)=\frac{1}{8}\int \frac{d^d{\bf
q}}{(2\pi)^d}\, \frac{1}{q^n(q+c)(q+k)}~,
\end{equation}
using a regularization cutoff $\Lambda>0$ at $d=3$. 
By definition, the regularized version of Eq. (\ref{d1}) is
\begin{equation}\label{d2}
J_n(k,c;\Lambda)=\frac{1}{8}\int_{-\Lambda}^{\Lambda} \frac{d^3
{\bf q}}{(2\pi)^3}\, \frac{1}{q^n(q+c)(q+k)}~,
\end{equation}
Employing the identity (valid for $A\neq B$)
\begin{equation}\label{d3}
\frac{1}{(q+A)(q+B)}=\frac{1}{B-A}\,\left[\,\frac{1}{q+A}-
\frac{1}{q+B}\,\right],
\end{equation}
one can recursively compute $J_n(k,c;\Lambda)$,
\begin{equation}\label{d4}
J_n(k,c;\Lambda)=\frac{1}{c-k}\,\left[\,J_{n-1}(k,0;\Lambda)-
J_{n-1}(0,c;\Lambda)\,\right]~,
\end{equation}
for $n\geq1$ from $J_0(k,c;\Lambda)$. This may be written from
(\ref{d2}) and (\ref{d3}) as
\begin{eqnarray}\label{d5}
J_0(k,c;\Lambda)&=&\frac{1}{8(c-k)}\,\int_{-\Lambda}^{\Lambda} 
\frac{d^3{\bf q}}{(2\pi)^3}\,\left[\, \frac{1}{q+k}-
\frac{1}{q+c}\,\right]\nonumber\\
&=&\frac{1}{16\pi^2}\Biggl[\,\Lambda+\frac{k^2}{c-k}\,{\rm ln}
\left(\frac{\Lambda}{k}\right)
-\frac{c^2}{c-k}\,{\rm ln}\left(\frac{\Lambda}{c}\right)
+O(\Lambda^{-1})\,\Biggr]~.
\end{eqnarray}
From (\ref{d4}) and (\ref{d5}) one gets
\begin{equation}\label{d6}
J_1(k,c;\Lambda)=\frac{1}{16\pi^2}\frac{1}{c-k}\,
\Biggl[\,-k\,{\rm ln}\left(\frac{\Lambda}{k}\right)+c\,{\rm ln}
\left(\frac{\Lambda}{c}\right)+O(\Lambda^{-1})\,\Biggr]~.
\end{equation}
Although the identity (\ref{d3}) is only valid for $A\neq B$, that
is $k\neq c$, the limit $k\to c$ of (\ref{d5}) and (\ref{d6}) is
well-defined,
\begin{eqnarray}\label{d7}
&&J_0(c,c;\Lambda)=\frac{1}{16\pi^2}\Biggl[\,
\Lambda-2c\,{\rm ln}\left(\frac{\Lambda}{c}\right)
+O(\Lambda^{-1})\,\Biggr]~,\nonumber\\
&&J_1(c,c;\Lambda)=\frac{1}{16\pi^2}\Biggl[\,{\rm ln}
\left(\frac{\Lambda}{c}\right)
+O(\Lambda^{-1})\,\Biggr]~.
\end{eqnarray}
Since $J_2(k,c)$ is regular in the UV one may take the limit
$\Lambda\to\infty$. The integral may be computed from (\ref{d1}) or
from (\ref{d4}), with the same result
\begin{equation}\label{d8}
J_2(k,c)=\frac{1}{16\pi^2}\,\frac{1}{c-k}\,
{\rm ln}\left(\frac{c}{k}\right)~.
\end{equation}
valid for $k\neq c$. In the case $k=c$ the integral gives
\begin{equation}\label{d9}
J_2(c,c)=\frac{1}{16\pi^2}\,\frac{1}{c}~.
\end{equation}

Finally, the functions $J'_n(k,c)=\partial J_n(k,c)/\partial k$
can be computed along the same lines, or using the explicit
formulas for $J_0(k,c;\Lambda)$, etc.~.

\bigskip\bigskip
%%%%%%%%%%%%%%%%%%%%%%%%%%%%%%%%%%%%%%%%%%%%%%%%%%%%%%%%%%%%%%%%%%
\renewcommand{\theequation}{C.\arabic{equation}}
\setcounter{equation}{0}
{\bf APPENDIX C}
\bigskip

In this Appendix we will show that  the joint contribution of
graphs (a)-(c) in Figure 1 to the one-loop connected four-point
function of the $\lambda\phi^4$ theory without the surface at
$z=0$ (bulk theory) can be written as in Eq. (\ref{94}).
Graph (a) may be obtained from (\ref{80}) upon replacing
the Robin propagator in (\ref{11}) by the bulk propagator
\begin{equation}\label{b2}
{\cal G}_B({\bf p};z,z^\prime)=\frac{1}{2p}\,e^{-p|z-z^\prime|}~,
\end{equation}
in the partial Fourier transform representation (\ref{7}).
It is enough to consider the
case $z_1=z_2=z_3=z_4=0$, since the bulk theory enjoy full
translational symmetry. Then after some manipulations it
is possible to break down $\tilde{\cal G}_B^{({\bf s})}
(\{{\bf p}_\ell\},\{0\})$ in two parts,
\begin{equation}\label{b3}
\tilde{\cal G}_B^{({\bf s})}(\{{\bf p}_\ell\},\{0\})=
\frac{\lambda^2}{2}\,\left[
\tilde{\cal G}_{1,B}^{({\bf s})}(\{{\bf p}_\ell\},\{0\})+
\tilde{\cal G}_{2,B}^{({\bf s})}(\{{\bf p}_\ell\},\{0\})\right]~.
\end{equation}

The term $\tilde{\cal G}_{1,B}^{({\bf s})}(\{{\bf p}_\ell\},\{0\})$ is
regular for $d<4$,
\begin{eqnarray}\label{b4}
&&\tilde{\cal G}_{1,B}^{({\bf s})}(\{{\bf p}_\ell\},\{0\})=
-\frac{1}{2}\,\prod_{\ell=1}^4
\frac{1}{2p_\ell}\,\int\frac{d^d{\bf p}}{(2\pi)^d}\,
\frac{1}{p~|{\bf p}-{\bf s}|}\,\int_0^\infty
dz\,dz^\prime\,e^{-(p_1+p_2)z-(p_3+p_4)z^\prime}\,
e^{-(p+|{\bf p}-{\bf s}|)(z+z^\prime)}\nonumber\\
&&\qquad\qquad\qquad=-\frac{1}{2}\,\prod_{\ell=1}^4
\frac{1}{2p_\ell}\,\int\frac{d^d{\bf p}}{(2\pi)^d}\,
\frac{1}{p~|{\bf p}-{\bf s}|~(p+|{\bf p}-{\bf s}|+p_1+p_2)~
(p+|{\bf p}-{\bf s}|+p_3+p_4)}~.
\end{eqnarray}
The other contribution contains a logarithmic divergence for
$d=3$,
\begin{eqnarray}\label{b5}
\tilde{\cal G}_{2,B}^{({\bf s})}(\{{\bf p}_\ell\},\{0\})&=&
\frac{1}{2}\,\prod_{\ell=1}^4 \frac{1}{2p_\ell}\,
\int\frac{d^d{\bf p}}{(2\pi)^d}\, \frac{1}{p~|{\bf p}-{\bf s}|}\,
\int_0^\infty dz\,dz^\prime
\,e^{-(p_1+p_2)z-(p_3+p_4)z^\prime}\,
e^{-(p+|{\bf p}-{\bf s}|)|z-z^\prime|}\nonumber\\
&=& \Lambda_{{\bf s}}\,\int_{-\infty}^\infty dz \,
\prod_{\ell=1}^4{\cal G}_B({\bf p}_\ell;0,z)~,
\end{eqnarray}
where $\Lambda_{{\bf s}}$ was defined in Eq.(\ref{89}), 
and we have used the identity
\begin{equation}\label{b6}
\int_{-\infty}^\infty dz \prod_{\ell=1}^4
{\cal G}_B({\bf p}_\ell;0,z)= \frac{2}{\sum_{k=1}^4 p_k}\,
\prod_{\ell=1}^4\frac{1}{2p_\ell}~.
\end{equation}
Altogether, Eq. (\ref{b3}) may be written as
\begin{equation}\label{b7}
\tilde{\cal G}_B^{({\bf s})}(\{{\bf p}_\ell\},\{0\})=
\lambda^2\left(\,
\Lambda_{{\bf s}}+\delta\Lambda_{{\bf s},B}\,\right)\,
\int_{-\infty}^\infty dz \prod_{\ell=1}^4 
{\cal G}_B({\bf p}_\ell;0,z)~,
\end{equation}
where $\delta\Lambda_{{\bf s},B}$ contains the regular
($d<4$) contributions given in Eq. (\ref{b4}). The other crossed
graphs (b) and (c) give similar contributions and one may write
\begin{equation}\label{b8}
\sum_{\beta={\bf s},{\bf t},{\bf u}}
\tilde{\cal G}_B^{(\beta)}(\{{\bf p}_\ell\},\{0\})=\lambda^2
\Biggl\{\,\sum_{\beta={\bf s},{\bf t},{\bf u}}\,\Lambda_{\beta}+
\sum_{\beta={\bf s},{\bf t},{\bf u}}\,\delta
\Lambda_{\beta,B}\,\Biggr\}\,
\int_{-\infty}^\infty dz \prod_{\ell=1}^4 
{\cal G}_B({\bf p}_\ell;0,z)~. 
\end{equation}
The more general case with $z_\ell\neq0$ preserves the structure
displayed in Eq. (\ref{b7}), with the same $\Lambda_{{\bf s}}$ 
on the RHS.

\newpage

%%%%%%%%%%%%%%%%%%%%%%%%%%%%%%%%%%%%%%%%%%%%%%%%%%%%%%%%%%%%%%%%%%%%

\begin{figure}[hbt]
\begin{center}
%\resizebox{!}{5cm}{\includegraphics{robinfig1.eps}}
\includegraphics*[scale=0.7, viewport = -120 -10 1300 500]{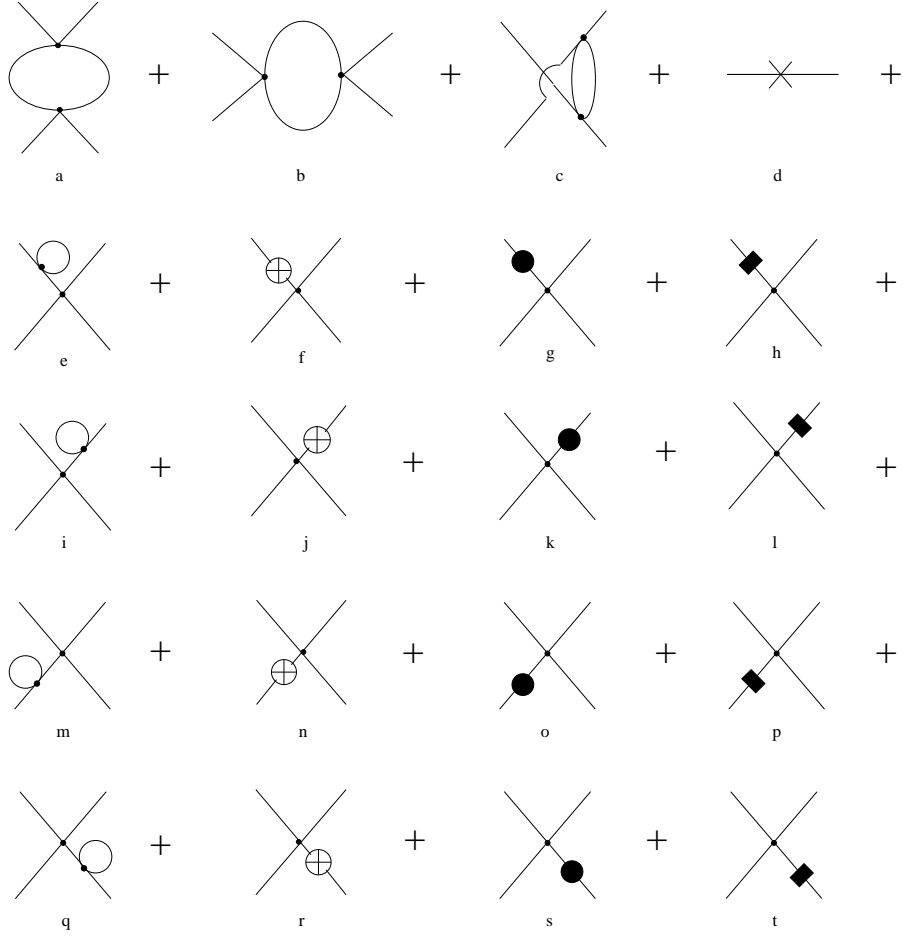}
\end{center}
\caption{Radiative corrections to the connected four-point
Green function at one-loop. Notation: cross in (d) denotes
$\delta\lambda_1$; crossed circle in (f) represent $\delta m_1$;
black circle in (g) represent the vertex (52);
and black box in (h) stand for the vertex (2).}%
\label{Fig1}
\end{figure}

\end{document}